\documentclass[a4paper]{article}
\usepackage{graphicx}

\newcommand{\EDW}{\textsc{Edelweiss}} %\mbox{
\title{\bf Latest results of the direct dark matter search with the EDELWEISS-2 experiment}
\date {}

\author{{\bf \normalsize Valentin Kozlov for the EDELWEISS collaboration}\\
{\small Karlsruhe Institute of Technology, Institut f\"ur Kernphysik, Postfach 3640, 76021 Karlsruhe, Germany}\\
{\small E-mail: {\bf Valentin.Kozlov@kit.edu}}}

\begin{document}

\maketitle

\begin{abstract}
{\EDW-2 is a Ge-bolometer experiment located in the underground laboratory Laboratoire Souterrain de Modane (LSM, France). For the second phase of the experiment, the collaboration has developed new cryogenic detectors with an improved background rejection (interleaved electrode design) \cite{edw09id}. A continuous operation of ten of these bolometers at LSM together with an active muon veto shielding has been achieved. First results based on an effective exposure of 144 kg$\cdot$d taken in 2009 have been published recently \cite{edw09dm}, the acquired data set has since then been doubled. The already published data correspond to an improvement in sensitivity of about 15 compared to \EDW-1 \cite{edw05dm}. We present and discuss the latest bolometer data including the identification of muon-induced background events and special measurements of muon-induced neutrons in LSM.}
\end{abstract}

{\small \it To appear in proceedings of PASCOS 2010, Valencia (Spain), 19-23 July 2010}

\section{Introduction}
A large number of cosmological observations, e.g. rotation curves of spiral galaxies, mass-to-light ratios of galaxy clusters measured by the use of gravitational lensing, anisotropy in cosmic microwave background (WMAP), can be consistently explained by introducing a non-visible matter in the Universe or so-called dark matter (DM). The required properties of a corresponding DM particle lead to the generic name WIMP, weakly interacting massive particle. An excellent candidate for this DM arises from the models of supersymmetry (SUSY), being the lightest SUSY particle, LSP. These particles can be searched for in a terrestrial detector by means of nuclear recoils originating from a WIMP-nucleon elastic scattering. The recoil energy, which has to be measured in such a direct detection experiment, can be calculated within the typically accepted models and it is in the range of 1 to 100~keV while the expected interaction rate is below 0.01 event per day and per kilogram of the target material. In order to be sensitive to this low count rate at such energies, experiments on direct DM search are commonly located deep underground to reduce the influence of cosmic rays, select low-radioactivity materials to be used, develop methods and the detectors themselves to have a very powerful background rejection, study  general background conditions of the experiment.

\section{EDELWEISS experiment}

\begin{figure}[!h]
    \begin{minipage}[c]{0.64\linewidth}
    \centering
        \includegraphics[width=0.98\textwidth]{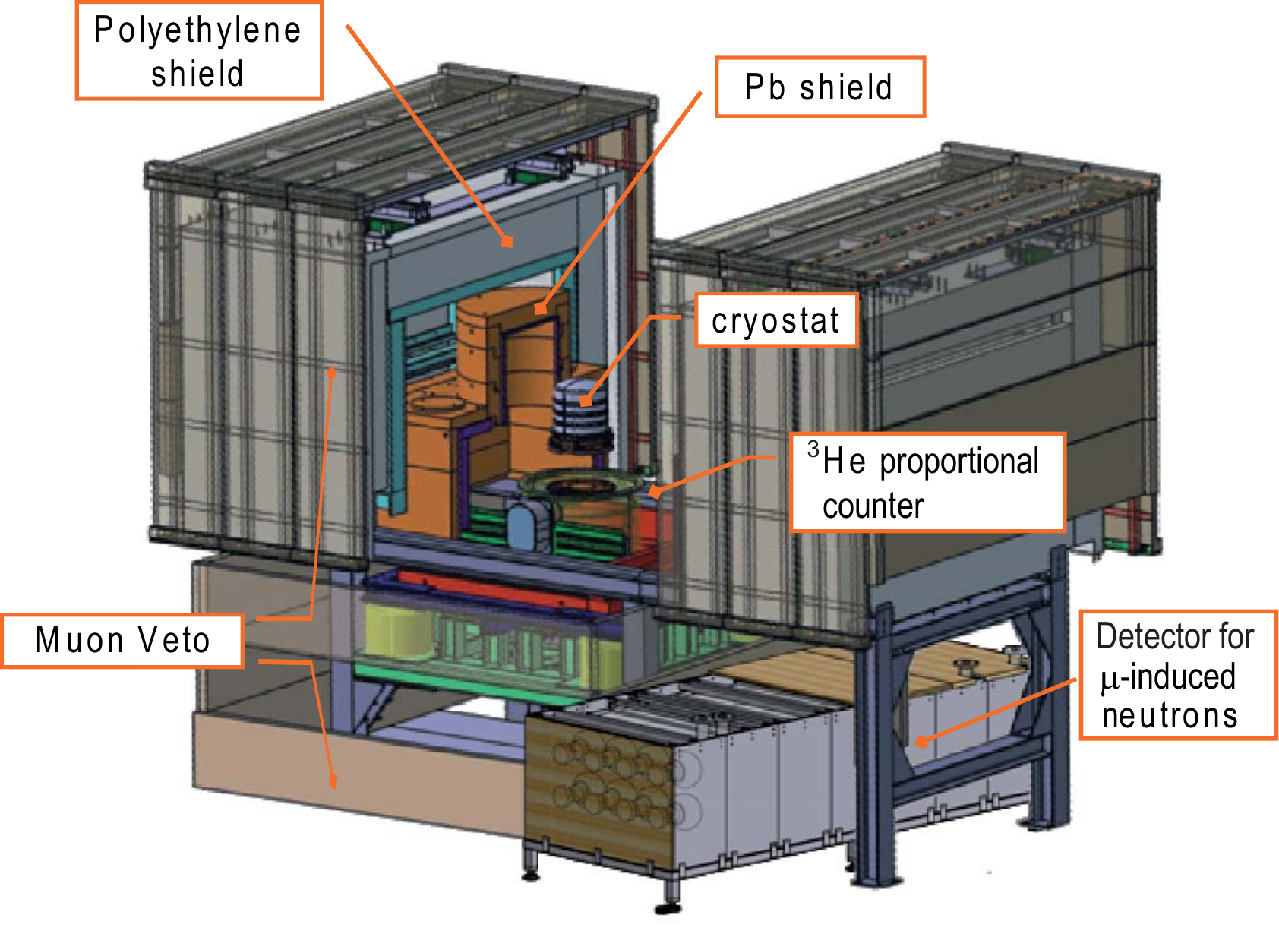}
    \caption{General layout of the \EDW-2 experimental set-up with additional $^3$He proportional counter for thermal neutrons and liquid scintillator detector to measure muon-induced neutrons.}
    \label{fig:edw-setup}
    \end{minipage}
    \hspace{4mm}
    \begin{minipage}[c]{0.33\linewidth}
        \centering
        \vspace{11mm}
        \includegraphics[width=0.9\textwidth]{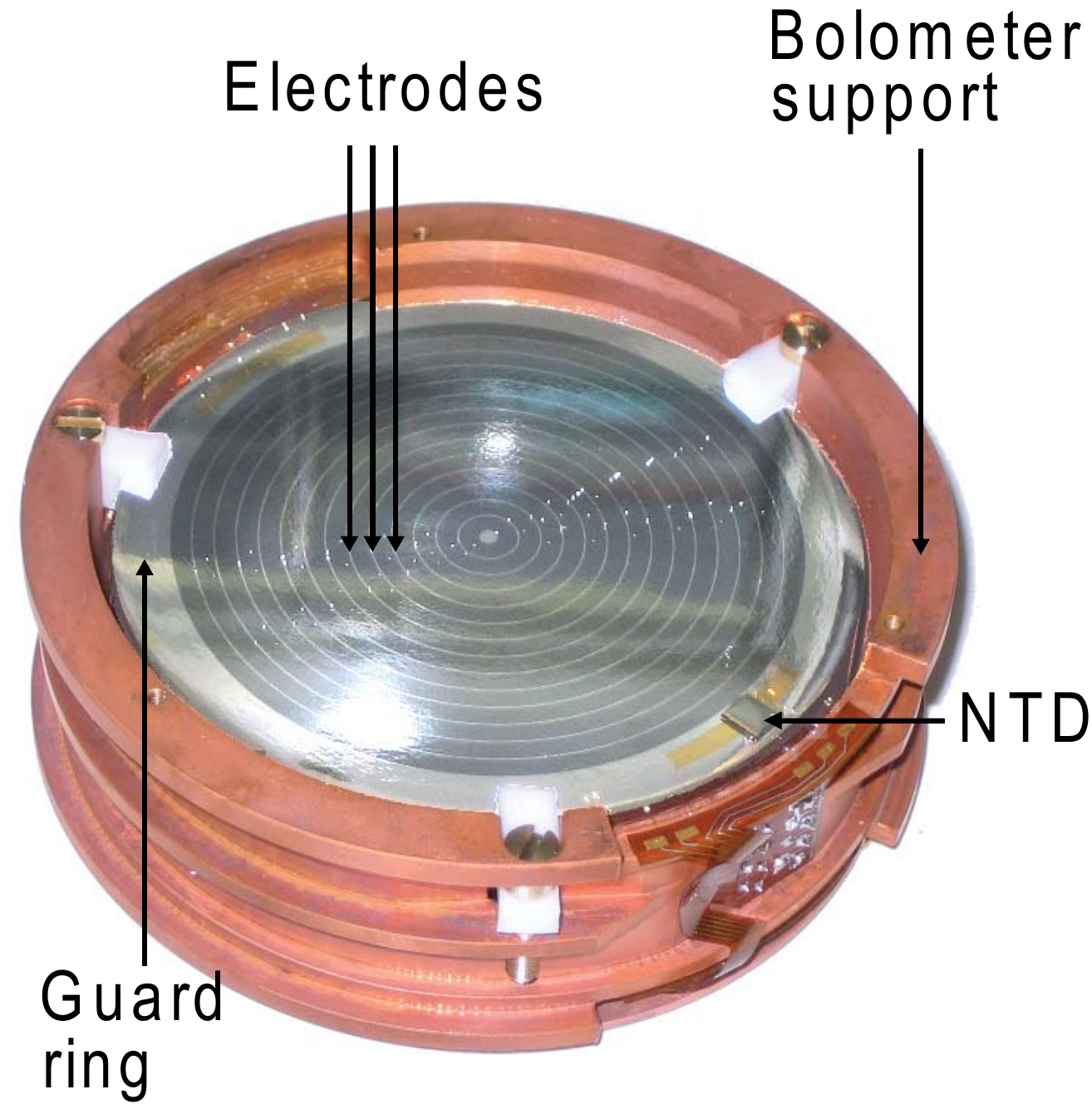}
        \vspace{11mm}
        \caption{400~g Ge bolometer with the interleaved electrode design.}
        \label{fig:edw-id}
    \end{minipage}
\end{figure}

%%%%%%%%%%%%%%%%%%%%%%%%%%%%
%\begin{figure}[!h]
%    \centering
%        \includegraphics[width=0.67\textwidth]{edw-setup-ppt.pdf}
%    \caption{General layout of the \EDW-2 experimental set-up. Position of $^3$He-gas detector and the neutron counter based on liquid scintillator are indicated as well.}
%    \label{fig:edw-setup}
%\end{figure}
%%%%%%%%%%%%%%%%%%%%%%%%%%%%

\EDW-2 is a direct DM search experiment (Fig.~\ref{fig:edw-setup}) located in the Laboratoire Souterrain de Modane (LSM), an underground lab in the Fr\'ejus French-Italian road tunnel in the Alps. This laboratory has a shielding of 4850~m.w.e., which reduces the muon flux down to about 5~$\mu$/m$^2$/day. Bolometers of pure natural Ge are used in the experiment both as the detectors and the target material. When these detectors are cooled down to about 20~mK, a deposited energy
of a particle can be measured by means of ionization and phonon excitation signals. Aluminum electrodes evaporated onto the top and the bottom surfaces of the Ge crystal are used to collect electrons and holes. The temperature increase is measured by the use of neutron-transmutation-doped germanium (NTD) as temperature sensor (Fig.~\ref{fig:edw-id}). The ratio of the two signals, so-called Q-value, is different for nuclear and electron recoils with nuclear recoils having Q$\sim$0.3 when normalized to Q=1 for electron recoils. This separation in Q allows a powerful $\gamma/\beta$-background rejection. A general overview of the set-up is shown in Fig.~\ref{fig:edw-setup}: The central part of the experiment is a dilution cryostat which can host up to 40~kg of detectors. A lead layer of 20~cm shields the bolometers against an external $\gamma/\beta$-background while 50~cm of polyethylene is used to moderate neutrons. A muon veto system ($>$98\% coverage) consisting of 100~m$^2$ of plastic scintillator to tag remaining cosmic muons completes the installation. In addition, a continuous control of the Rn level is performed near the cryostat, a $^3$He-gas detector is installed inside the shields to monitor the thermal neutron flux. A neutron counter system based on 1~m$^3$ of Gd-loaded liquid scintillator \cite{edw10nc} and coupled electronically with the muon veto allows to study muon-induced neutron background. The scientific goal of \EDW-2 is to reach a sensitivity of 10$^{-8}$~pb for the WIMP-nucleon spin-independent (SI) cross-section in 2011 and, with an upgrade of the set-up, to achieve 5$\times$10$^{-9}$~pb by 2013.

\section{Bolometers with interleaved electrodes}
Although the background conditions in \EDW-2 are improved in comparison to \EDW-1 by a factor of 3 for $\gamma$'s and a factor of 2 in contamination of $^{210}$Pb, bolometers with higher rejection power than used in \EDW-1 had to be developed in order to be able to achieve the aimed sensitivity for DM search. This especially concerns the rejection of so-called surface events \cite{edw09id}, which have Q-values in the region-of-interest (ROI) due to an incomplete charge collection near the detector surface. The \EDW\ collaboration thus has developed and successfully tested new bolometers with interleaved electrode design (Fig.~\ref{fig:edw-id})\cite{edw09id}. A main principle of these detectors, named inter-digitized electrode detectors, or ID's, is to have concentric ring electrodes for charge collection instead of one planar electrode. By tuning the electric potentials applied on the neighboring electrodes the electrical field is modified near the surface such that depending on an interaction position, i.e. in the bulk or near the surface, a created charge is accumulated either on a set of electrodes with the highest potential, called  \textit{fiducial} electrodes, or in addition on low potential electrodes, called correspondently \textit{veto} electrodes. This allows to reliably reject the near-surface events \cite{edw09id}. A first bolometer of this kind was tested for a first time in November 2007 having 200\,g mass with 50\% as fiducial volume, in January 2009 ten of the 400~g detectors were installed in LSM for the direct DM search.

\section{One year of WIMP search with ID detectors}

\begin{figure}[!h]
    \begin{minipage}[c]{0.5\linewidth}
        \centering
        \includegraphics[width=0.90\textwidth]{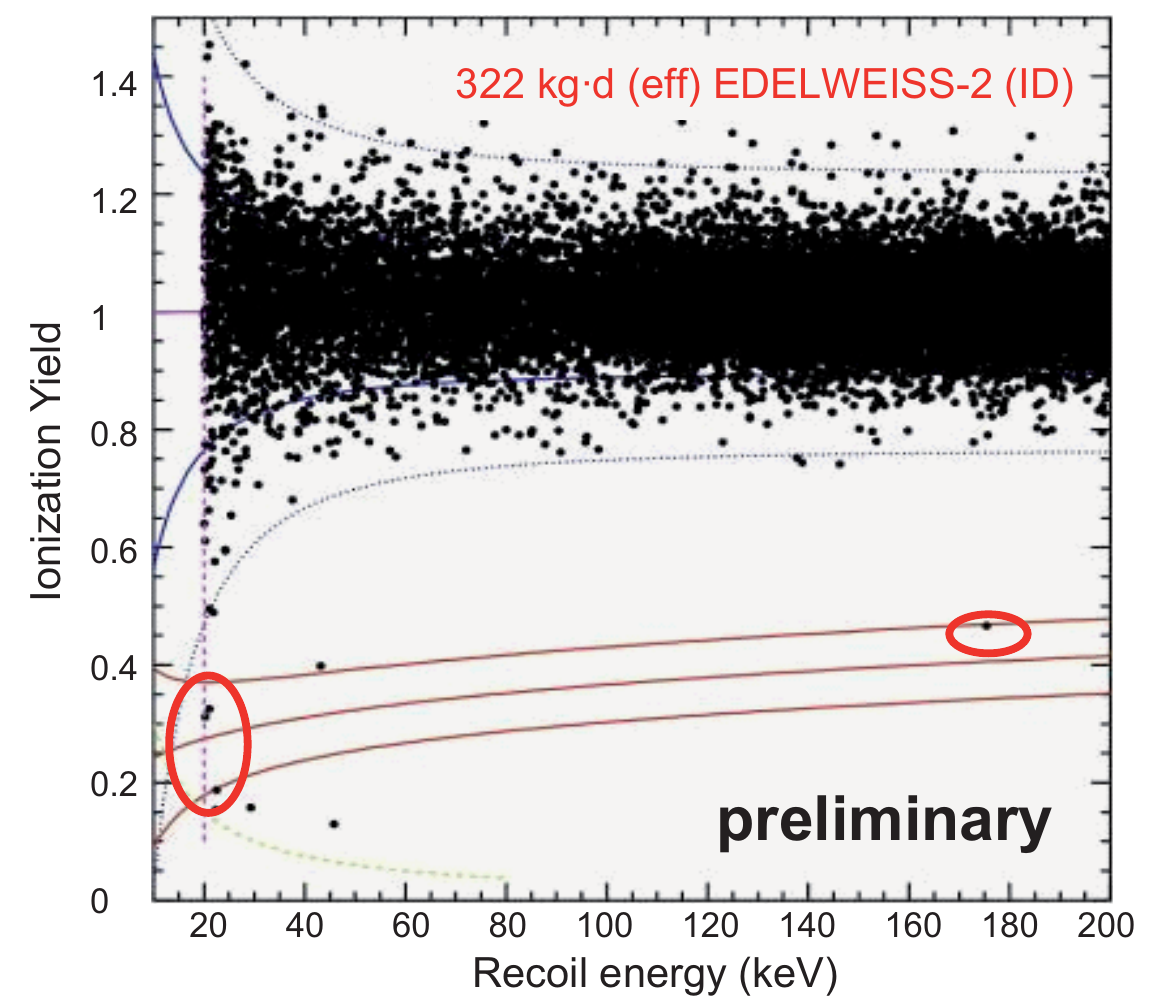}
    \end{minipage}
    \begin{minipage}[c]{0.5\linewidth}
        \centering
        \includegraphics[width=0.9\textwidth]{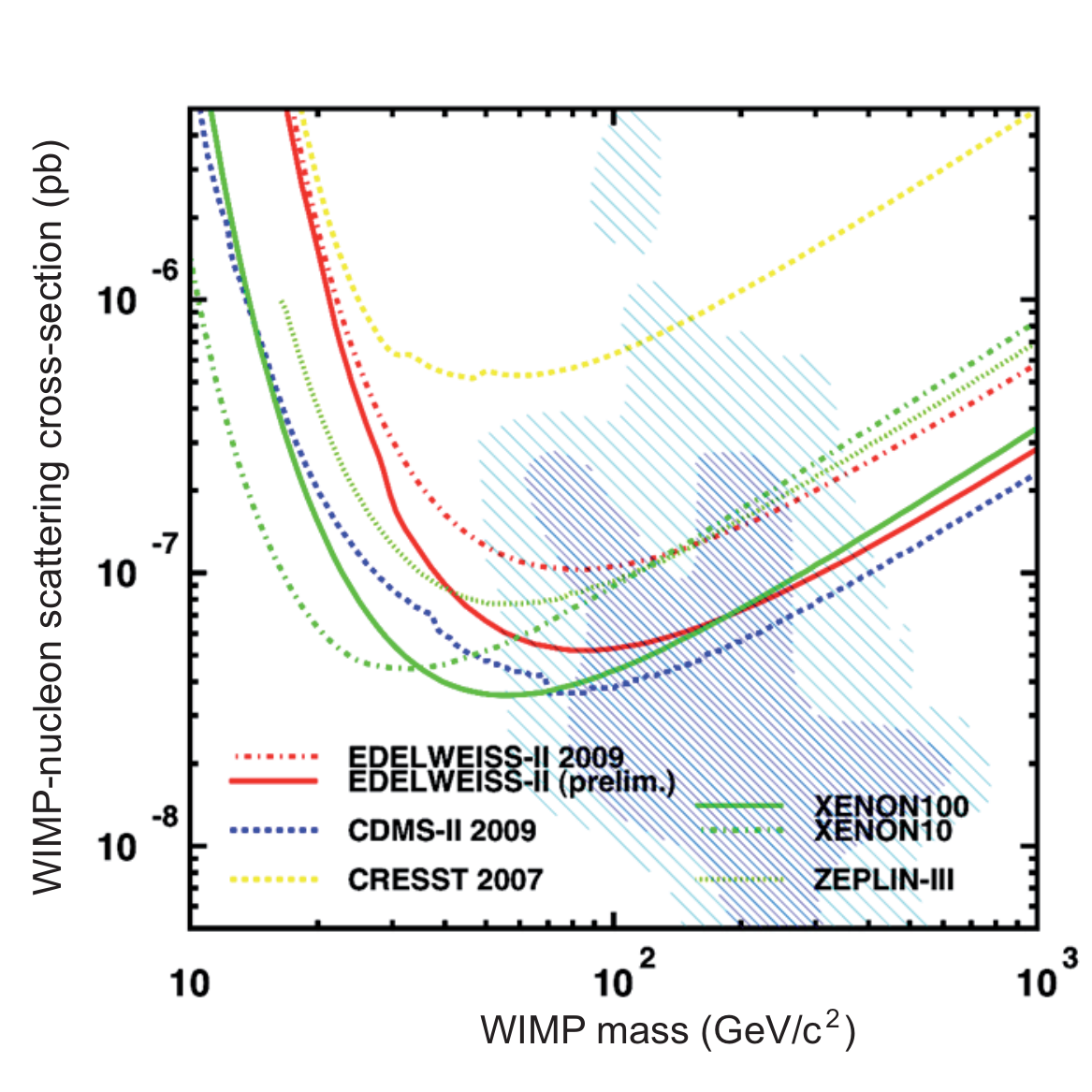}
    \end{minipage}
    \caption{Recent results of the Edelweiss-2 experiment: one-year data obtained in the physics run with new ID bolometers, 4 events passed all the cuts are encircled (\textit{left}); the upper limits on the WIMP-nucleon spin-independent cross-section as a function of WIMP mass (\textit{right}).}
    \label{fig:edw-results}
\end{figure}

After commissioning of the new detectors, they were continuously operated starting from April 2009 to May 2010 constituting 418 days in total, out of which 322 days were under stable running conditions and 305 days have been considered for the WIMP search, the remaining 17~days being used for various calibrations. During a physics run, a regular energy calibration with $^{133}$Ba source was performed as well as every day regeneration of the detectors in order to avoid the formation of space charge. After the first 6 months of operation, the data were analyzed offline with 2 independent analysis chains \cite{edw09dm}. Data of 9 out of 10 detectors were used in the analysis. One detector had one deficient guard and one missing veto electrode and in the absence of reliable proof that information from these two electrodes can be efficiently compensated by the remaining channels, this detector was left out of the analysis. The WIMP search threshold was fixed \textit{a priori} to 20~keV in order to ensure that the data set is compatible with a nearly full efficiency for the chosen energy threshold. The data with appropriate baseline resolution and online heat threshold were selected on an hour-by-hour basis. This requirement reduced the effective exposure by about 20\%. A reduced $\chi^2$ cut was applied on the pulse fit for heat and fiducial signals of each event. The efficiency of this cut is controlled with the $\gamma$-ray background and its typical value is 97\%. The fiducial mass was measured individually for each detector using the low-energy peaks ($\sim$10~keV) associated with the cosmogenic activation of germanium, resulting in a uniform contamination by the isotopes of $^{65}$Zn and $^{68}$Ge. An average fiducial mass of 166$\pm$6~g was deduced for 400~g detectors and a conservative value of 160~g is used in the calculation of the total exposure. Both analyses agreed in the final result of the six-month run: one nuclear recoil candidate was observed in an effective exposure of 144~kg$\cdot$d. Two further events were in coincidence with either an other bolometer or the muon veto and thus excluded as a WIMP candidate. This result was intepreted in terms of limits on the spin-independent cross-section of WIMP scattering off nucleons, e.g. a cross-section of 1.0$\times$10$^{-7}$~pb is excluded at 90\%~CL for a WIMP mass of 80~GeV/c$^2$ (Fig.~\ref{fig:edw-results}(\textit{right}, \EDW-II 2009)). This result corresponds to an improvement in sensitivity of about 15 compared to \EDW-1 \cite{edw05dm} and demonstrates the very high rejection capabilities of ID detectors in an actual WIMP search. Detailed description of the six-month analysis can be found in Ref.~\cite{edw09dm}.

\noindent In May 2010 the WIMP search run was stopped and the full data set was analyzed by blindly applying the same cuts as for the first six months. Preliminary results of one of the two analysis chains (second one is still ongoing as for the time of writing) are shown in Fig.~\ref{fig:edw-results}. In an effective exposure of 322~kg$\cdot$d there are in total 4 events located in the nuclear recoil band which passed all the cuts: three are close to the threshold and one is at 175~keV (Fig.~\ref{fig:edw-results}(\textit{left})). Using an unbiased method to determine the energy window of optimal sensitivity \cite{yellin02} leads to an upper limit on the WIMP-nucleon elastic scattering cross-section of 5$\times$10$^{-8}$~pb (90\%~CL) for a WIMP mass of 80~GeV/c$^2$ (Fig.~\ref{fig:edw-results}(\textit{right}, \EDW-II (prelim.))). The fact that there is no event between 23 and 175~keV results in an almost linear improvement of sensitivity for WIMP masses above 50~GeV, i.e. an improvement of $\sim$2 compared to the data in the 6-months analysis published earlier \cite{edw09dm}. The 4 observed events are to be compared with expectations for various background sources discussed in the following.

\section{Background studies}
The present estimate of an upper limit of the total background of known origin is 1.6 events (90\%~CL) in the corresponding exposure of 322~kg$\cdot$d. The different sources being studied at present are the residual gamma, surface events caused mainly  by the remaining beta activity, neutron background from radioactive decays and the muon-induced neutrons. The fact that in the WIMP run 3 events are observed very close to the threshold points towards the need to better understand the detector performance and more detailed studies to determine the acceptance of events in this region. 

Regular $\gamma$-calibrations with $^{133}$Ba source allow to study the rejection power of all 10 detectors together against the $\gamma$-background. The number of events found in the nuclear recoil band in ROI during these calibrations compared to the total number of $\gamma$-events in the detector bulk yields the $\gamma$-rejection factor, which is then used to estimate the corresponding contribution in the WIMP search run. One can still improve the $\gamma$-rejection power by optimizing the polarization voltages of the bolometers, implement 2 NTD sensors for heat measurement, introduce different electrode combinations for segmentation. A faster readout electronics can help to avoid pile-up events. The number of unrejected surface events is estimated by multiplying the number of observed low-ionization yield events in the WIMP search run before the rejection of surface events by the rejection power measured previously in Ref.~\cite{edw09id}. The surface event background of ID detectors can be further decreased with a better surface treatment, optimization of the mass-to-surface ratio and by improving an energy resolution. The contribution of neutrons from radioactive decays has been simulated using GEANT3 in Ref.~\cite{chabert04} and studied in Ref.~\cite{chardin03,rozov10}. The presence of U/Th traces in the lead shield was recently measured as well as the efficiency of \EDW\ neutron shields was checked with an external strong AmBe source ($\sim$10$^5$ neutrons/s). The number of neutrons from interactions of untagged muons is estimated from the number of observed coincidences between the bolometers and the muon veto, multiplied by the muon veto efficiency. The effect of neutrons can be further reduced by an additional polyethylene layer between the lead shield and the cryostat. Comprehensive GEANT4 simulations were developed \cite{horn07} within the collaboration in order to understand the muon-induced background in detail. In the simulations it was shown, for instance, that the muon-induced bolometer events with an energy deposition of E$\geq$50~keV are dominantly the electron recoils while events with E$\leq$50~keV are neutron recoils. Analysis of real data measured in coincidence between the muon veto and bolometers highlights this fact experimentally \cite{chantelauze07,chantelauze09}. However, the analysis of data suffers from the low coincidence rate, e.g. only several events are detected in 100 live days. This is why a special neutron counter based on a liquid Gd-loaded scintillator was designed and installed. It routinely acquires data since end of 2008 and the count rate of muon-induced neutrons in the order of 1~neutron/day has been achieved \cite{edw10nc}.

\section{Conclusion and outlook}
\EDW-2 experiment successfully runs since end of 2007 aiming to reach sensitivity on WIMP-nucleon SI cross-section of 10$^{-8}$~pb in 2011 and 5$\times$10$^{-9}$~pb by 2013. New bolometers with the interleaved electrode design have been developed for direct DM search. Ten of these detectors were continuously operated in LSM during 418~days together with an active muon veto shielding. Preliminary data analysis of the full effective exposure of 322~kg$\cdot$d shows an improvement in sensitivity of about 30 compared to \EDW-1, e.g. a cross-section of 5$\times$10$^{-8}$~pb is excluded at 90\%~CL for a WIMP mass of 80~GeV/c$^2$. The detector development continues towards further improvement and 4 of new \textit{fully interdigitized electrode} bolometers, each with a total mass of 800~g and 80\% fiducial volume, were installed in LSM in July 2010. This development together with the detailed studies of the background conditions in LSM, in particular, muon induced neutrons give us the confidence to reach aimed sensitivity  and put a good base for a DM experiment of next generation, e.g. EURECA \cite{kraus07}, a 1-ton cryogenic detector array.

\section*{Acknowledgements}
This work is supported in part by the German Research Foundation (DFG) through its collaborative research center SFB-TR27 ("Neutrinos and Beyond"), by the French Agence Nationale pour la Recherche and the Russian Foundation for Basic Research (grant No. 07-02-00355-a).

%\section*{References}

\end{document}